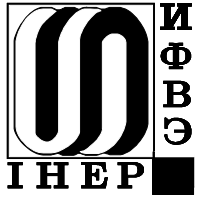

A.G.Afonin,[1] V.T.Baranov,[1] S.Bellucci,[2] S.A.Belov,[1] S.Bini,[2]
V.N.Gorlov,[1] G. Giannini,[2] A.D.Ermolaev,[1] I.V.Ivanova,[1]
D.M.Krylov,[1] V.A.Maisheev,[1] D.A.Savin,[1] E.A.Syshchikov,[1]
V.I.Terekhov,[1] V.N.Chepegin,[1] Yu.A.Chesnokov,[1]
P.N.Chirkov,[1] I.A.Yazynin.[1]

1 - SRC IHEP, Protvino, Moscow region, 142281, Russia

2 – LNF INFN, Frascati, 00044, Italy

# Study of Beam Extraction and Collimation in the U-70 Synchrotron with Various Crystal Devices

**Abstract**

Afonin A.G. et al. Study of Beam Extraction and Collimation in the U-70 Synchrotron with Various Crystal Devices: IHEP Preprint 2010–12. − Protvino, 2010. – p. 11, figs. 12, tables 1, refs.: 10.

Novel crystal technique - array of bent strips and a veer-type reflector, based on thin straight plates - have been used for research of extraction and collimation a circulating beam in the U-70 accelerator at the energy 50 GeV and 1.3 GeV. It is shown, that new devices can effectively steer a beam in a wide energy range. For protons with energy 50 GeV efficiency of extraction and collimation about 90 % has been achieved which is record for this method. Reduction of particle losses in 2-3 times was observed also in accelerator at application of different crystals in comparison with the usual one-stage collimation scheme of beam with a steel absorber.

**Аннотация**

Афонин А.Г. и др. Исследование вывода и коллимации пучка в синхротроне У-70 с помощью различных кристаллических дефлекторов: Препринт ИФВЭ 2010–12. − Протвино, 2010. – 11 с., 12 рис., 1 табл., библиогр.: 10.

Новая кристаллическая техника – массив изогнутых полосок и отражатель веерного типа, основанный на тонких прямых пластинах – были использованы для исследования вывода и коллимации циркулирующего пучка в ускорителе при энергии 50 ГэВ и 1.3 ГэВ. Показано, что новые устройства могут эффективно управлять пучком в широком интервале энергий. Для протонов с энергией 50 ГэВ была достигнута рекордная для этого метода эффективность вывода и коллимации около 90%. Наблюдалось также уменьшение потерь частиц в 2-3 раза на ускорителе за поглотителем при применении разных кристаллов в сравнении с обычной одноступенчатой схемой коллимации пучка стальным поглотителем.

# 1. Introduction

The phenomenon of deflection of a charged particle beam in a bent crystal is well investigated and successfully applied for beam extraction in high-energy accelerators, at energies of about 10 GeV and higher (see for example Ref. [1, 2, 3] ). However, the task of bending and extraction of charged particles with energies below 1 GeV presents a big practical interest, for example for the production of ultra stable beams of low emittance for medical and biological applications. There exists a big experimental problem in steering such energy beams, which is connected with the small size of the bent crystal samples. The efficiency of particles deflection is determined by the ratio of the critical channeling angle $\theta_c$ to the beam divergence j and drops exponentially with the crystal length L:

$$\text{Eff} \sim (\theta_c / \varphi) \times \exp(-L / L_d),$$

where the characteristic parameter $L_d$, called dechanneling length, is relatively small for low energy. For example, for E = 500 MeV, we have $\theta_c$ = 0.24 mrad and $L_d$ = 0.4 mm. With usual channeling bent crystals (about 1 mm in length) only 10% efficiency was achieved for the deflection of sub – GeV energy particles [4] in beam line.

Still the big problems arise in a task of extraction of a circulating beam from the ring accelerator as in addition the significant cross-section sizes of a crystal exceeding its length here are required. Thus the bend angle of a crystal should be more than 1 mrad that the deflected beam was well separated from circulating one. Potentially suitable tools in this case can be the bent quasimozaic crystals such as in [5], or thin straight crystals [6, 7], but in both these cases it is necessary to increase a deflection angle of particles in some times.

# 2. Crystal Devices

In this article we propose a novel crystal technique, which can effectively work in a wide energy range and is especially perspective for low energy below 1 GeV.

The first option is based on use of array of shot bent channeling crystals (Fig. 1) with sub – millimeter length (special thin silicon wafers about 100 micron thickness were used for the production of such samples). Thus the bend of array occurs also, as a bend of the single well investigated silicon strip [8].

The second option is based on the reflection of particles on very thin straight crystal plates with thickness, which is equal to an odd number of half-lengths of channeling oscillation waves $L = (2n+1)/2 \times \lambda$, where $\lambda = \pi d/\theta_c$, $d = 2.3 A^0$ – interplanar distance in silicon. It means, for example, that the optimum length of a crystal should be 10 microns for particles with energy 50 GeV. The reflection angle in one silicon plate should be equal to twice the critical angle $\theta_c = (2U_o/pv)^{1/2}$, where $U_o \sim 20$ eV – is the value of the potential of planar channel in silicon, p, v - a momentum and speed of incident particle.. For the enhancement of the deflection angle, a few aligned plates placed like a veer are foreseen (Fig. 2).

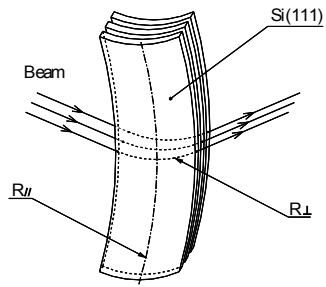

Fig. 1. The array of bent silicon strips for beam deflection due to channeling.

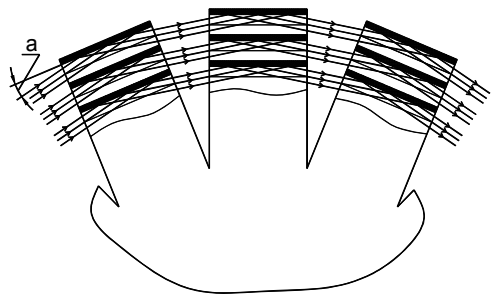

Fig. 2. Veer-type reflector for bending of particle beam with use of thin straight crystals. Reflection of trajectories of particles from nuclear planes is schematically shown.

For an optimum deflection of a beam in this design each following crystal is unwrapped on angle $2\theta_c$. Then the total bend angle of beam can reach the value $2\theta_c \times N$, where где N – is the amount of crystal plates. If the unwrap of a fan and thickness of plates are not optimal, the lower bend of particles occurs, and this picture is more difficult for interpretation. On Fig. 3 for understanding of the process Monte Carlo calculations for unitary passage of a beam through a fan are submitted at its different parameters.

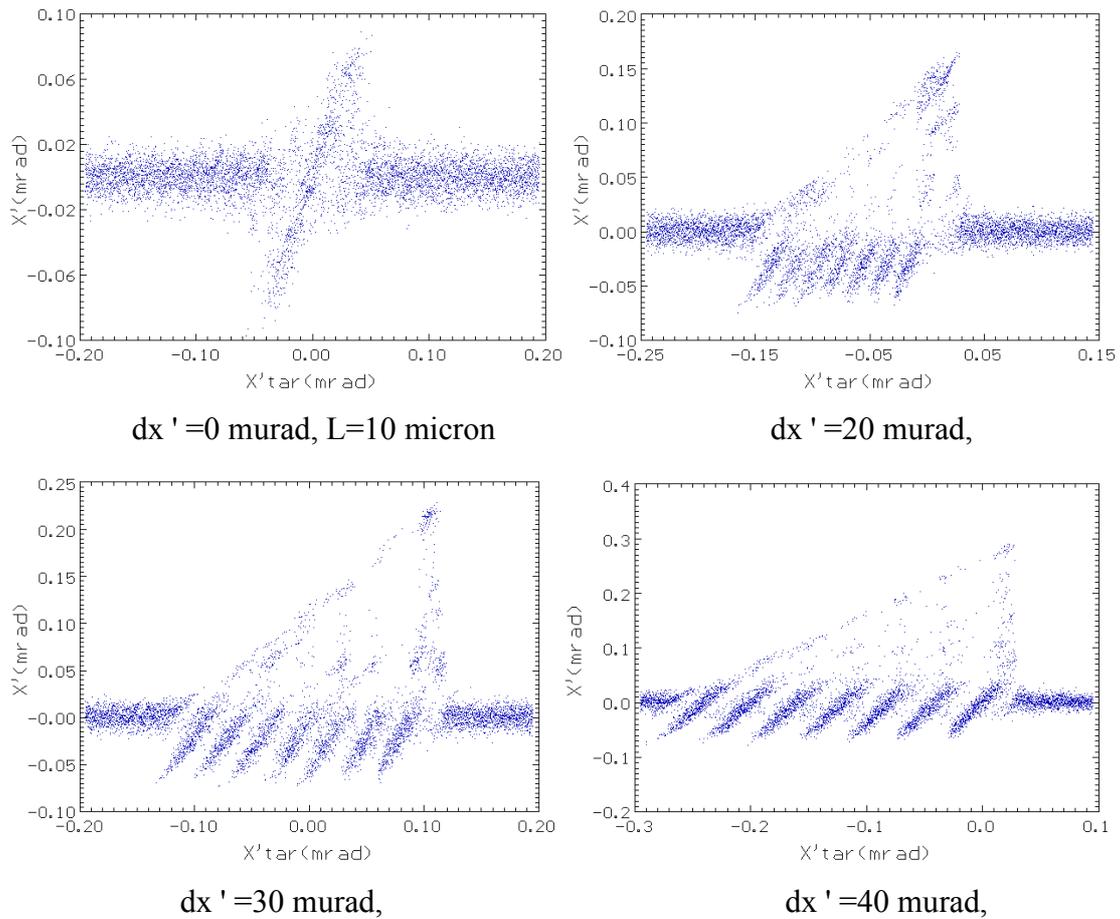

dx ' =0 murad, L=10 micron      dx ' =20 murad,

dx ' =30 murad,      dx ' =40 murad,

Fig. 3. Distribution of protons with energy 50 GeV on scattering angles after passage of a crystal fan from seven plates depending on its angular orientation with respect to the beam. The parameter dx ' means a turn of the next plates of a fan in microradians.

Three different devices have been prepared for accelerator experiment: a usual crystal strip (the technology is described in [8]), a crystal array ( Fig. 1), and assembly of veer - type ( Fig. 2). The realized schemes of crystal devices are shown in photos Fig. 4.

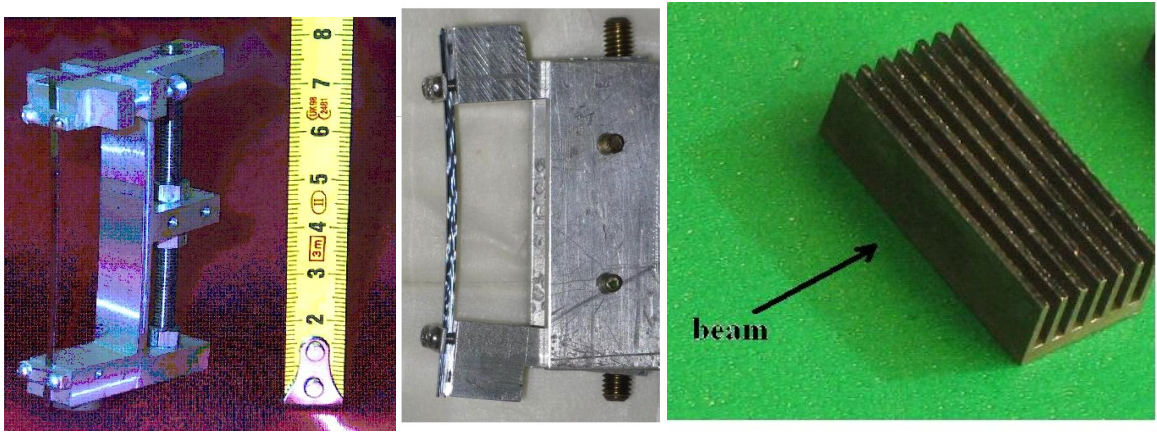

Fig. 4. The realized schemes of devices (left to right): the bent strip, array of strips, a fan of plates.

Parameters of crystals are submitted in the table.

| Crystal type | Length along the beam, mm | Transversal Dimension, mm | Deflection angle, mrad | Collimation Efficiency, % | | Channeling peak Efficiency, % | |
|---|---|---|---|---|---|---|---|
| | | | | 50 GeV | 1.3 GeV | 50 GeV | 1.3 Gev |
| Simple strip | 1 | 0.3 | 1.0 | 91±2.5 | 30±2.5 | 86±2.5 | ~ 0 |
| Crystal array | 0.9 | 7×0.2 | 1.1 | 77±2.5 | 40±2.5 | 72±2.5 | ~ 20 |
| Veer | 7×0.25 | 3 | ~7×0.05 | 82±2.5 | 35±2.5 | 70±2.5 | ~ 5 |

## 3. Description of Accelerator Experiment

Experiment was performed during two runs of U-70 operation in 2009 and 2010. In experiment 3 crystals in high-vacuum goniometers were serially entered in a circulating accelerated beam as shown in Fig. 5. This angle of bending is sufficient to separate the circulating and deflected (by the crystal) beams in space. The beam deflection effect due to channeling was measured by secondary emission detector (SEM), located in vacuum chamber of an accelerator near to circulating beam. Parameters of the accelerator are described in [9]. Details of the experimental equipment are shown on Fig. 6.

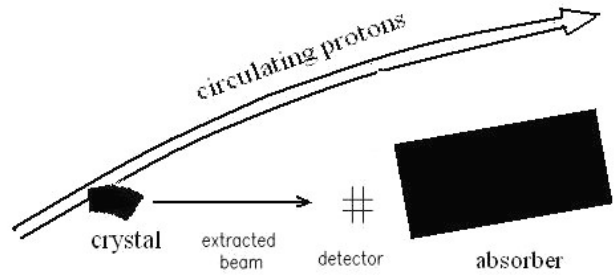

Fig. 5. The scheme of experiment on beam extraction and collimation by crystals.

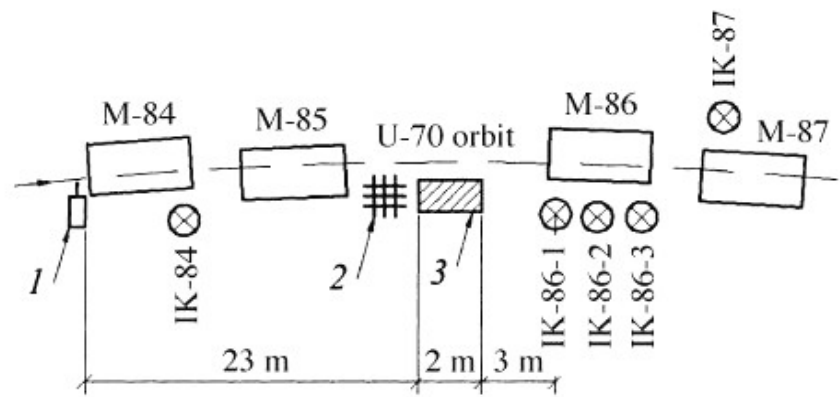

Fig. 6. An arrangement of the equipment in the U-70 accelerator: 1 - crystal station, 2 - profilometer, 3 - an absorber, IK – ionization chambers (loss monitors 1-5), M - magnetic blocks of the accelerator.

For carrying out of experiment the new automated system of diagnostics of the beam, including 5 monitors of losses based on ionization chambers and 2 plane stationary profilometer has been created on basis of the secondary emission, established directly ahead of an absorber in vacuum volume and recording parameters of the deserted beam in horizontal and vertical planes.

The system of data taking and acquisition is based on use of an industrial computer with trunk PCI. In the tunnel of the accelerator two skeletons of preliminary electronics for integration of signals from electrodes of profilometer are placed. The first skeleton contains 24 measuring channels for registration of horizontal distribution of density of a beam and 8 channels for integration of signals from loss monitors. The second skeleton with 32 measuring channels is intended for 32 channel vertical profilometer. In each skeleton signals from integrators are multiplexed and on the common coaxial cables act on 2 inputs of 8-channel ADC, integrated in an industrial computer. On other inputs of ADC other necessary auxiliary information, somehow acts: intensity of a beam before its dump on an absorber, the voltage values proportional to currents of bump-magnets, to positions of crystals, etc. The software for data taking and acquisition has been based on application of platform LabVIEW.

## 4. Results of Measurements

Measurements were carried out at two energies of the accelerated beam of protons: 50 GeV and 1.3 GeV (kinetic energy).

First measurements have been lead at energy of protons 50 GeV. Fig. 7 illustrates the beneficial effect of crystals when used as primary element of system concerning a beam (the beam was brought to a crystal with the help of slowly increasing bump). It shows beam profiles in the radial direction 23 meters downstream of the crystal as measured on the entry face of the absorber. Four cases are reported. First, the end face of amorphous absorber is used as primary target while the crystal is kept outside of the beam envelope. As expected, the beam profile is peaked at the absorber edge. The second, third and fourth cases correspond to aligned crystals: single strip, silicon array and veer – type. In these cases crystals channel most of the incoming particles into the depth of the absorber. Deep brought of particles improves collimation, or can be used for extraction of circulating particles from the accelerator.

For definition of collimation efficiency calibration of the detector with the help of fast kicker - magnet has been lead. In Fig. 8 the sums of signals of the detector are shown when the beam deflected by different crystals in comparison with effect of one-turn brought of a beam by fast kicker (in this last case whole beam with 6 mm size gets in the aperture of the detector). Measured collimation efficiencies for three mentioned crystals are equal 91, 77 and 82 %. The amount of particles in channeling peak is accordingly equal 86, 72 and 70 %. That is all crystals well deflected 50 GeV proton beam. The received data are in agreement with computer simulation with use of program SCRAPER [10] which is taking into account multi-turn character of movement of particles in real structure of the accelerator and repeated their interaction with a crystal.

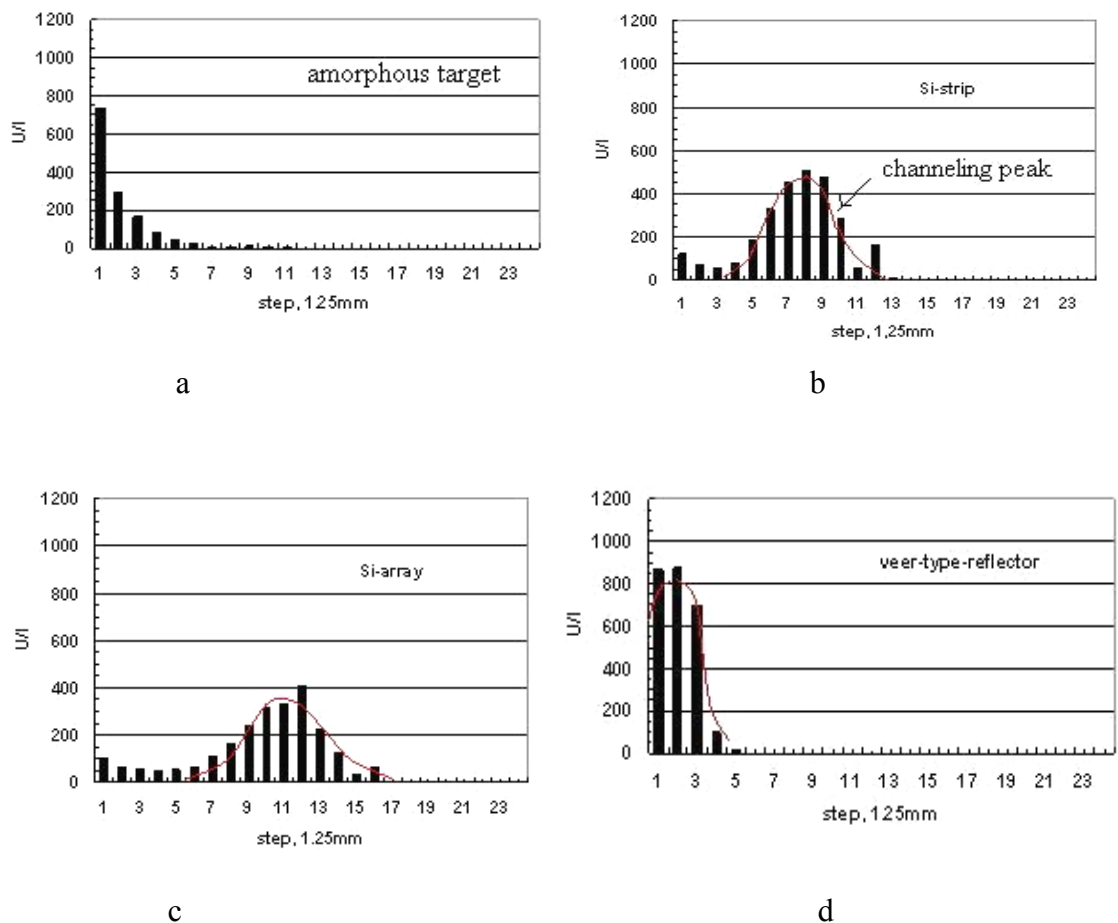

Fig. 7. Beam profiles on entry face of the absorber in different cases.

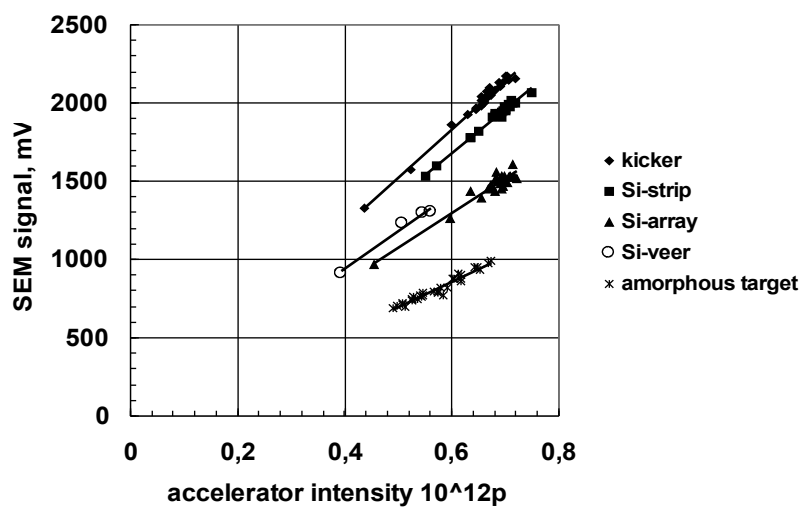

Fig. 8. Comparison of the sum of profilometer signals at work of different crystals with effect of particle brought towards absorber by kicker.

On Fig. 9 orientation dependences of collimation and channeling efficiencies are shown for different crystals.

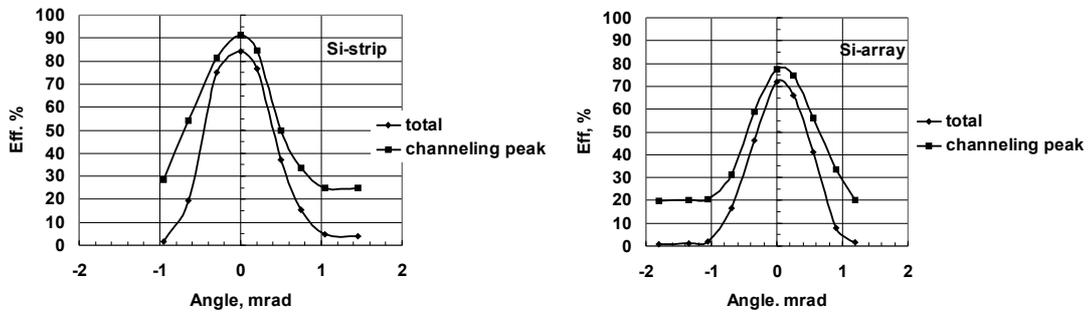

Fig. 9. Collimation efficiency depending on angular orientation of a crystal with respect to the beam (at the left - for a strip, on the right - for the array of strips).

On Fig. 10 are shown orientation curves of particle losses, measured on signals of 5 ionization chambers located in a vicinity of an absorber.

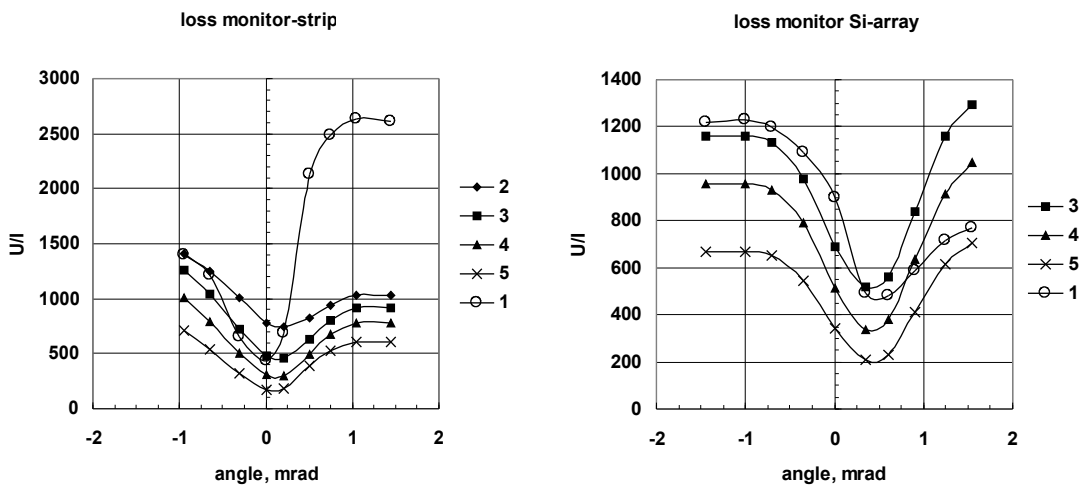

Fig. 10. Orientation dependences of loss monitors for a strip crystal (at the left) and array of strips (on the right).

Particle losses at optimum alignment of crystals decrease in 2-3 times in comparison with disoriented crystals that corresponds to calculation. Approximately in as much time intensity of muon torch behind an absorber far from the accelerator should decrease that is the important factor at achievement of high intensity of beam of circulating protons in the accelerator. On Fig. 11 effects of reduction of losses of particles on the accelerator behind an absorber are shown at application of different crystals in comparison with the usual one-stage scheme of beam collimation by a steel absorber.

At energy of a beam 1.3 GeV results of efficiency are much lower (see the table). The best result, channeling peak of about 20 %, has shown array from seven thin strip crystals. The big loss of efficiency is explained by non-optimal tuning of circulating beam towards the crystal by bump-magnet. At low energy because of sizable beam, about 50 mm, there is a drift of an incident angle about a half milliradian that should be removed at prompting a beam by high-frequency noise (this work is planned). In Fig. 12 the profile of 1.3 GeV beam is shown deflected by silicon array. The fraction of channeling peak is allocated by a thick line (channeling peak is well separated from a circulating beam and approximately corresponds to efficiency of a possible beam extraction from the accelerator).

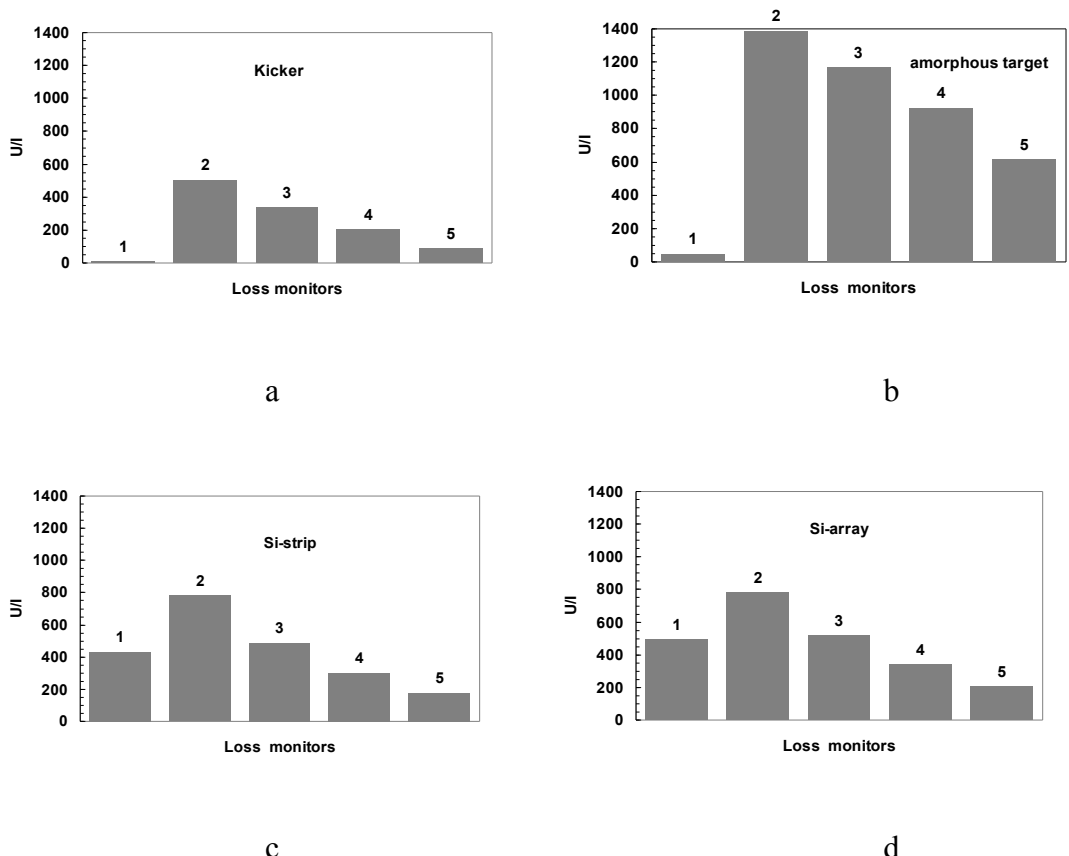

Fig. 11. a - losses of particles behind an absorber at brought of a beam by kicker magnet (collimation efficiency is 100 % in this case), b - losses at usual collimation by edge of absorber, c - application of a strip – type crystal, d - application of array of strips.

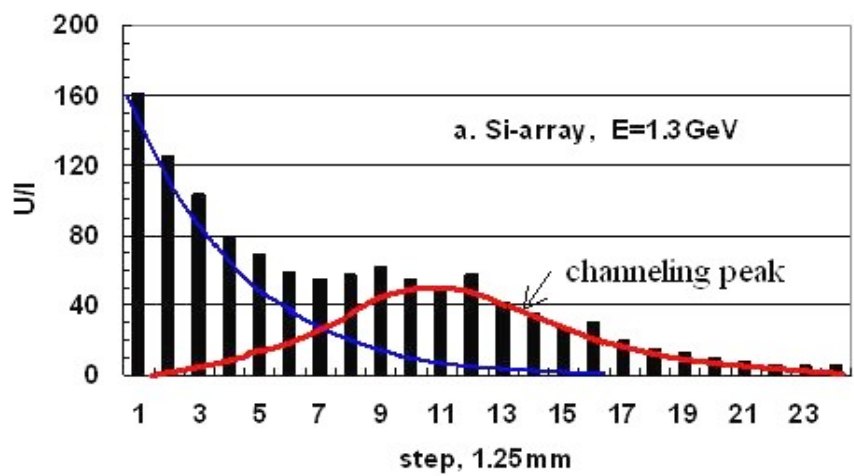

Fig. 12. The 1.3 GeV proton beam profile at the absorber entry deflected by array of silicon strips.

## 5. Conclusion

Thus, it is shown, that the created crystal devices can work in a wide range of energy and have prospects for the organization of low energy medical beams on U-70 accelerator. For optimization of crystal devices for low energy experiments in a test area of laboratory LNF are planned where accessible beam of particles with 50-700 MeV energies exist [4].

Reduction of particle losses in 2-3 times on the accelerator was observed also at application of different crystals in comparison with the usual one-stage scheme of beam collimation by steel absorber. Approximately in as much time intensity of muon torch behind an absorber far from the accelerator should decrease that is the important factor for achievement of higher intensity of a circulating beam in U-70.

Work is supported by IHEP Directorate, State corporation Rosatom (the contract № 4e.45.03.09, 1047), the grant of the Russian Funding for Basic Research 08-02-01453-a, and also the grant 09-02-92431-KE of joint project of RFBR - Consortium EINSTEIN (Italy).